\documentclass[conference,10pt]{IEEEtran}
\usepackage{epsfig,rotating,setspace,latexsym,amsmath,epsf,amssymb,amsfonts,bm,subfigure,epstopdf,cite,authblk,bbm,mathtools,amsthm,algorithm,color,systeme,mathtools,amsmath,derivative,bigints}
\usepackage[
top=1.86 cm,
bottom = 2.75cm,
left   = 1.6cm,
right  = 1.9cm]{geometry}
\usepackage[utf8]{inputenc}
\usepackage[T1]{fontenc}
\usepackage[thinc]{esdiff}
\usepackage[noend]{algpseudocode}
\theoremstyle{plain}
\newtheorem{theorem}{Theorem}
\newtheorem{lemma}{Lemma}
\newtheorem{corollary}{Corollary}

\newtheorem{remark}{Remark}
\newenvironment{Proof}[1]{\medskip\par\noindent{\bf Proof:\,}\,#1}{{\mbox{\,$\blacksquare$}\par}}

\algrenewcommand\algorithmicforall{\textbf{foreach}}
\algrenewcommand\algorithmicindent{.8em}
\usepackage{tabularx}

\DeclarePairedDelimiter{\ceil}{\lceil}{\rceil}

\IEEEoverridecommandlockouts
\allowdisplaybreaks

\begin{document}

\title{Tracking and Assigning Jobs to a Markov Machine }

\author{Subhankar Banerjee \qquad Sennur Ulukus\\
\normalsize Department of Electrical and Computer Engineering\\
\normalsize University of Maryland, College Park, MD 20742\\
\normalsize \emph{sbanerje@umd.edu} \qquad \emph{ulukus@umd.edu} }

\maketitle

\begin{abstract}
We consider a time-slotted communication system with a machine, a cloud server, and a sampler. Job requests from the users are queued on the server to be completed by the machine. The machine has two states, namely, a busy state and a free state. The server can assign a job to the machine in a first-in-first-served manner. If the machine is free, it completes the job request from the server; otherwise, it drops the request. Upon dropping a job request, the server is penalized. When the machine is in the free state, the machine can get into the busy state with an internal job. When the server does not assign a job request to the machine, the state of the machine evolves as a symmetric Markov chain. If the machine successfully accepts the job request from the server, the state of the machine goes to the busy state and follows a different dynamics compared to the dynamics when the machine goes to the busy state due to an internal job. The sampler samples the state of the machine and sends it to the server via an error-free channel. Thus, the server can estimate the state of the machine, upon receiving an update from the source. If the machine is in the free state but the estimated state at the server is busy, the sampler pays a cost, as the sampler aims to deliver the state information of the machine to the server in a timely manner. We incorporate the concept of the age of incorrect information to model the cost of the sampler. We aim to find an optimal sampling policy such that the cost of the sampler plus the penalty on the machine gets minimized. We formulate this problem in a Markov decision process framework and find how an optimal policy changes with several associated parameters. We show that a threshold policy is optimal for this problem. We show a necessary and sufficient condition for a threshold policy to be optimal. Finally, we find the optimal threshold without bounding the state space.
\end{abstract}

\section{Introduction}
In the recent literature, much work has been done on tracking the state of a source that follows different dynamics, see e.g., \cite{sun2019sampling, sun2019sampling2, ornee2021sampling, kriouile2022minimizing, bountrogiannis2024age, salimnejad2024version, kam2020age, champati2022detecting, chen2021scheduling, joshi2021minimization, chen2024minimizing, kriouile2021minimizing, maatouk2020age, kam2018towards,bastopcu2021timely,cosandal2024modeling,akar2024timely,cosandal2024aoii}. Specifically, the works in \cite{kriouile2022minimizing, bountrogiannis2024age, salimnejad2024version, kam2020age, champati2022detecting, chen2021scheduling, joshi2021minimization, chen2024minimizing, kriouile2021minimizing, maatouk2020age, kam2018towards} consider tracking the state of Markov sources. Most of the works consider a system where a sampler samples the states of the source and delivers that to a monitor. Based on the system requirement, we see that different works use different penalty functions to optimize the sampling policy, e.g., \cite{salimnejad2024version, delfani2024semantics, luo2024minimizing}. One of the prevalent metrics that has been used in the literature for this tracking problem is the age of incorrect information, which was first introduced in \cite{maatouk2020age}. 

After the introduction of the age of incorrect information, many works consider this metric in the context of tracking. However, these papers only consider the tracking problem of the source. It is not always clear what we will achieve by tracking the source, i.e., a clear motivation behind the purpose of tracking a source is usually lacking. In this work, we consider a Markov machine, which has two states, namely a busy state and a free state. Whenever the machine is in the free state it can go to the busy state to perform some internal job, and after finishing that job, it can go back to the free state, following the dynamics of a symmetric Markov chain.

\begin{figure}[t]
    \centerline{\includegraphics[width = 1\columnwidth]{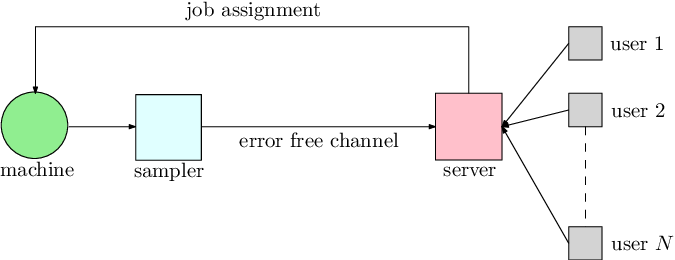}}
    \caption{System model: A machine has two states, namely busy and free. A sampler samples the state of the machine and sends it to the server. Job requests from users are queued on the server to be completed by the machine. The server assigns jobs to the machine in a first-in-first-served manner.}
    \label{fig:1}
    \vspace*{-0.4cm}
\end{figure}

The system also includes a cloud server and a sampler. The cloud server receives job requests from multiple external users. In this work, we assume that the job requests from all external users are of the same nature, thus the server does not distinguish between users while assigning a job to the machine. The server aims to track the state of the machine, as whenever the machine is in the free state, the server submits a job request to it.  However, when the server assigns a job to the machine and the machine is busy, the machine has to drop the job request. Whenever the machine drops a job request, it pays a penalty. Thus, tracking the state of the machine in a timely manner is crucial for this application. The goal of the sampler is to sample and transmit the state of the machine such that the server has the correct information about the state of the machine. When the machine is in the free state and can serve a job request from the server, however, the estimated state of the machine in the server is busy, i.e., the server does not assign a job to the machine, the sampler incurs a linearly increasing penalty, similar to the age of incorrect information, as there is a waste of the resource.  Fig.~\ref{fig:1} shows the different components of the considered system model.

When the server assigns a job to the machine and the machine is in the free state, it accepts the job and goes to the busy state immediately. Then, it follows a different dynamics than the original dynamics to complete the job and comes out of the busy state. To the best of our knowledge, in the context of tracking with the age of incorrect information penalty, this is the first work that considers assigning a job while tracking a Markov machine and changing its dynamics after accepting the job. We will discuss how this change in dynamics affects the sampling policy. Note that, this problem is different than tracking a non-homogeneous Markov chain, as the change in dynamics occurs based on the sampling decisions. We formulate the sampling problem as a Markov decision process (MDP). We then show how different parameters affect an optimal sampling policy, namely, the transition probabilities of the Markov machine, the slope of the age of incorrect information, and the penalty corresponding to dropping a job. We show that the optimal sampling policy has a threshold structure. There exist two real numbers, such that an optimal action for the sampler is to sample the state of the machine and transmit it to the server if the age of incorrect information lies between these two numbers. We obtain these two numbers by necessary and sufficient conditions for sampling to be optimal, which we also study in this work. Finally, we find the optimal threshold without bounding the state space and without running any iterative algorithm. Due to space limitations, we do not provide all the proofs, which will be provided in a journal version of this paper. Here, we provide some representative proofs.

\begin{figure}[t]
    \centerline{\includegraphics[width = 1\columnwidth]{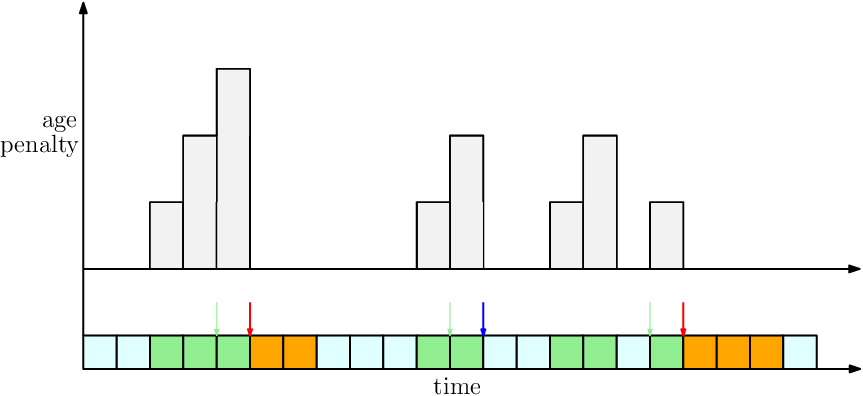}}
    \caption{Pictorial representation of the dynamics of the Markov machine and the age of incorrect information penalty. The cyan boxes represent the busy state of the Markov machine when it is not serving any job request from the server, the green boxes represent the free state of the Markov machine, and the orange boxes represent the busy state of the Markov machine when it is serving a job request from the server. The green arrows represent the sampling instances, the red arrows represent an assignment of a job to the machine and the machine accepts that job, and the blue arrow represents an assignment of a job to the server, however, it does not accept the job as the machine has transitioned into the busy (internally) state, and thus pays the penalty $p$. We use $S=2$, for the evolution of age of incorrect information.}
    \label{fig:2}
    \vspace*{-0.4cm}
\end{figure}

\section{System Model and Problem Formulation}
We consider a fully time-synchronized system. At time $t$, we denote the state of the machine with $m(t)$, $m(t)=\{0,1\}$, where $0$ denotes the busy state and $1$ denotes the free state. The communication channel between the sampler and the server is error-free, and it takes one time-slot to transmit a status sample. Based on the received sample, the server estimates the state of the machine. Following the literature, we assume that the server estimates the state of the machine as the most recently received sampled state. We denote the estimated state at time $t$, with $\hat{m}(t)$.

The transmission of a sample starts at the beginning of a time slot and finishes at the end of a time slot. The job request from the server to the machine is instantaneous, and the server always makes a job request at the beginning of a time slot. 

If the machine is free and the server does not submit a job request to the machine, then it follows the dynamics of a symmetric Markov chain, with the transition probability being $q>0$. If the machine is free and the server submits a job request to the machine, then the state of the machine goes to the busy state. In that case, the machine comes out of the busy state with probability $q_{1}>0$, in every time slot. In other words, the service time of the machine for a job assigned by the server is geometrically distributed, with the probability of success being $q_{1}$.  
 
\begin{figure}[t]
    \centerline{\includegraphics[width = 0.7\columnwidth]{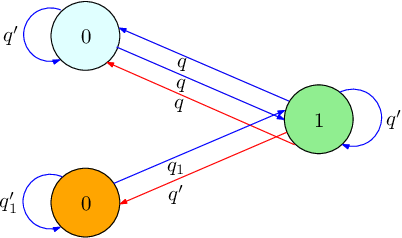}}
    \caption{Dynamics of the states of the Markov machine. The cyan circle corresponds to the busy state when the machine is not serving any job request from the server,  the orange circle corresponds to the busy state when the machine is serving a job request, and the green circle corresponds to the free state. The blue arrow corresponds to action $a=0$, while the red arrow corresponds to action $a=1$. Here, we assume that $q'=1-q$ and $q_{1}'=1-q_{1}$.}
    \label{fig:4}
    \vspace*{-0.4cm}
\end{figure}
 
Under any dynamics, it takes a one-time slot for the machine to change the state. For example, if the machine is free at the beginning of a time slot, and the server does not submit a job request, then the state of the machine becomes busy at the end of that time slot with probability $q$. We assume that the server always has a job available in its queue. We also assume that whenever the estimated state of the machine at the server is free, the server assigns a job to the machine. Due to the one-slot delay in transmission, the estimated state can be free, however, the actual state of the machine may be busy. In this situation, the machine drops the job request and pays a penalty $p>0$. At time $t$, we define the penalty on the sampler as $v(t)=S\cdot(t-u(t))$, where $S>0$, and $u(t)$ is defined as,
\begin{align}
     u(t)= \sup_{t'\leq t} \big\{\hat{m}(t)=m(t')\big\}.
\end{align}
Note that, $v(t)$ is similar to the age of incorrect information in the literature \cite{maatouk2020age}. A pictorial representation of the age of incorrect information dynamic together with the Markov machine dynamics is given in Fig.~\ref{fig:2}.

Note that, when $q$ is large enough, the probability that the machine is busy, but the estimated state of the machine at the server is free, is high. This increases the machine's penalty for dropping a job request. Thus, for large $q$, intuitively, the sampler should sample the machine for a large age of incorrect information. In other words, the sampler aims to reduce the penalty for dropping a job request at the expense of increasing the age of incorrect information, for large $q$. Similarly, for low $q$, the sampler aims to reduce the age of incorrect information, at the expense of the penalty for dropping a job request. This shows that there is a trade-off between the penalty corresponding to dropping a job request and the penalty regarding not sampling even when the machine is in the free state.

At time $t$, we denote the job request denial by the machine with $P(t)\in\{0,1\}$, here $0$ implies that the machine accepts the request if there is any, and $1$ implies otherwise. Note that, if at the beginning of time $t$, $m(t)=\hat{m}(t)$, the sampler does not sample and transmit the state of the machine. Also note that $v(t+1)$, is always $0$, if at time $t$, the sampler samples and transmits the state of the machine to the server. Under a sampling policy $\pi$, at time $t$, we denote the action of the sampler with $\pi(t)\in{\{0,1\}}$, where $\pi(t)=1$ implies that the sampler samples the state of the machine and transmits it to the server, and $\pi(t)=0$ implies otherwise. The dynamics of the machine are pictorially represented in Fig.~\ref{fig:4}. We are interested in the following problem,
\begin{align}\label{eq:2}
    \inf_{\pi\in\Pi} \limsup_{T\rightarrow \infty} \frac{1}{T} \sum_{t=1}^{T} \mathbb{E}_{\pi}[v(t) + p P(t)].  
\end{align}

\section{Optimal Policy}
We formulate the problem in the MDP setting.
We represent the state of the system with a vector $s=(v,b)$, where $v$ is the age of incorrect information and $b$ can take values $0$ or $1$. When $b=0$, it implies that the machine is not serving any job request from the server, and $b=1$ implies otherwise. The previous section shows that when $b=1$, the corresponding $v$ is always $0$. The state space is denoted by $\mathcal{S}$. Thus, formally,
\begin{align}\label{eq:3}
    \mathcal{S} = \{(v,0):v\in\mathcal{N}\}\cup \{0,1\}.
\end{align}
The action of the sampler is denoted by $a\in\{0,1\}$, which is similar to the $\pi(t)$. From the previous section, we see that when the $v=0$, the sampler does not sample. At time $t$, if $v\neq 0$ and the sampler samples the state, then at the beginning of slot $t+1$, the server sends a job request to the machine. However, at the end of slot $t$, the state of the machine changes with probability $q$, and thus it pays a cost $p$ at time $t+1$, due to the action taken in time $t$. Thus, we denote the cost of the MDP as,
\begin{align}
    C(s,a) = Sv +  a q  p. 
\end{align}
Under action $a$, we denote the transition from state $s$ to state $s'$, with $P_{a}(s,s')$. For the simplicity of the presentation, we define some of the states below,
\begin{align}
    &s_{1}=(v,0), \quad s_{2} = (v+1,0), \quad s_{3} = (1,0), \nonumber\\ 
    & s_{4}=(0,1), \quad \!\!\ s_{5}=(0,0).
\end{align}
Next, we present all possible non-zero transition probabilities,
\begin{align}\label{eq:6}
    &P_{1}(s_{1},s_{5}) = q, \quad P_{1}(s_{1},s_{4}) = 1-q, \quad P_{0}(s_{1},s_{2}) =1-q, \nonumber\\ 
    & P_{0}(s_{1},s_{5})=q, \quad P_{0}({s_{5},s_{5}}) = 1-q, \quad P_{0}(s_{5}, s_{3}) = q, \nonumber\\
    & P_{0}(s_{4},s_{4}) = 1-q_{1}, \quad P_{0}(s_{4},s_{3})= q_{1}.
\end{align}
In the MDP formulation, we can write (\ref{eq:2}) as,
\begin{align}\label{eq:7}
      \inf_{\pi\in\Pi} \limsup_{T\rightarrow \infty} \frac{1}{T} \sum_{t=1}^{T} \mathbb{E}_{\pi}\big[C(s(t),a(t))\big].
\end{align}
For self-containedment, we introduce some of the well-known results in the MDP literature. For $\alpha>0$, and an initial state $s'\in{\mathcal{S}}$, the discounted cost for a policy $\pi$ is defined as,
\begin{align}
    V_{\alpha}^{\pi}(s') = \sum_{t=1}^{\infty}\alpha^{t}\mathbb{E}_{\pi}[C(s(t),a(t))|s'],
\end{align}
and the $\alpha$ discounted value function is defined as,
\begin{align}
    V_{\alpha}(s) = \inf_{\pi\in{\Pi}} V_{\alpha}^{\pi}(s).
\end{align}
From \cite{sennott1989average}, we know that if for all $s\in{\mathcal{S}}$, $V_{\alpha}(s)$ is finite, then the discounted value function satisfies the Bellman equation i.e., $V_{\alpha}(s) = \min_{a\in{\{0,1\}}} V_{\alpha}(s;a)$, where,
\begin{align}
    V_{\alpha}(s;a) = C(s,a) + \alpha \sum_{s'\in{\mathcal{S}}} P_{a}(s,s')V_{\alpha}(s').
\end{align}
Now, we define the following iteration,
\begin{align}
    &V_{\alpha,n}(s) = \min_{a\in{\{0,1\}}} V_{\alpha,n}(s;a),\label{eq:11}\\&
     V_{\alpha,n}(s;a) = C(s,a) + \alpha \sum_{s'\in{\mathcal{S}}} P_{a}(s,s')V_{\alpha,n-1}(s').\label{eq:12}
\end{align}
From \cite{ross2014introduction} we know that,
\begin{align}\label{eq:13}
    V_{\alpha}(s) = \lim_{n\rightarrow\infty} V_{\alpha,n}(s).
\end{align}
Now, we introduce the relative value function \cite{bertsekas2012dynamic}, which we will use later to prove some of the results in this paper. First, we define a bounded state space,
\begin{align}
    S_{K} = \{(v,0):v\in\mathcal{N}, v\leq K\}\cup \{0,1\}.
\end{align}
Let us consider the following iteration for $s\in{\mathcal{S}_{K}}$, 
\begin{align}\label{eq:nn15}
    V_{n}(s) = &\min_{a\in{\{0,1\}}}\big\{C(s,a)+\sum_{s'\in{\mathcal{S}}}P_{a}(s,s')V_{n-1}(s') \nonumber\\
    &\qquad \qquad - V_{n-1}(s_{5})\big\},
\end{align}
where $V_{0}(s)=0$, $s\in{\mathcal{S}_{K}}$. Following  \cite[Proposition~4.3.2]{bertsekas2012dynamic}, we say that the limit, $\lim_{n\rightarrow\infty} V_{n}(s)$ exists, which we denote with $V(s)$. Again from \cite[Proposition.~4.3.2]{bertsekas2012dynamic}, we say that the following relation holds,
\begin{align}\label{eq:nn16}
    V(s) = &\min_{a\in{\{0,1\}}}\big\{C(s,a)+\sum_{s'\in{\mathcal{S}}}P_{a}(s,s')V(s') \nonumber\\
    &\qquad \qquad  - V(s_{5})\big\}.
\end{align}
From \cite[Proposition~4.2.1]{bertsekas2012dynamic}, we say that the policy which satisfies the right-hand side of (\ref{eq:nn16}) is an optimal policy, and $V(s_{5})$ is the optimal cost. Note that, we do not use the iteration of (\ref{eq:nn15}) to find an optimal policy, we only use (\ref{eq:nn15}) and (\ref{eq:nn16}) to prove some of the properties of an optimal policy. Thus, we can take $K$ as large as possible, without worrying about the computational complexity to find an optimal policy.

We say that a policy $\pi$ is stationary if its action only depends on the current state of the system and is independent of time. Similar to the techniques studied in \cite{banerjee2024preempt, banerjee2023re, banerjee2024wiopt, banerjeeitw}, we can show that there exists a stationary policy that is optimal for the problem considered in (\ref{eq:7}). Similar to \cite{banerjee2024preempt, banerjee2023re, banerjee2024wiopt, banerjeeitw}, we say that most of the properties that are true for the optimal policy corresponding to the discounted problem also hold for the average cost problem, i.e., for (\ref{eq:7}). Thus, to prove any property of an optimal policy for (\ref{eq:7}), that satisfies the above-mentioned relation, we prove the property for the optimal discounted cost problem. If the above-mentioned relation does not hold for certain properties, we will mention it specifically. We say a stationary policy $\pi$ is a threshold policy, if for a state $s=(v,0)$,
$\pi$ selects action $a=1$, then for any state $s'=(v+x,0)$, $\pi$ has to select action $a=1$, for any positive $x$. In the next theorem, we study the optimality of threshold policies. Before that, we present the next lemma, which studies the monotonic behavior of the discounted value function with $v$.
\begin{lemma}\label{lemma:1}
    For $\alpha>0$, the discounted value function is a monotonically increasing function of $v$.
\end{lemma}
\begin{Proof}
    We prove this with mathematical induction on $n$ in the iteration (\ref{eq:12}), and then from  (\ref{eq:13}), we have this lemma. For $n=0$, the statement is obvious. Let us assume that $V_{\alpha,n-1}(s)$, where $s=(v,0)$ is an increasing function of $v$.  Now,
    \begin{align}
        V_{\alpha,n}(s;1) = & Sv + qp + \alpha\big(q V_{\alpha,n-1}(s_{5}) \nonumber\\ &+ (1-q) V_{\alpha,n-1}(s_{4})\big),\label{eq:14} \\ V_{\alpha,n}(s;0)= & Sv + \alpha \big(q V_{\alpha,n-1}(s_{5}) \nonumber\\&+ (1-q) V_{\alpha,n-1}(s_{2})\big).\label{eq:15}
    \end{align}
    From (\ref{eq:14}), (\ref{eq:15}), and induction step $(n-1)$, $V_{\alpha,n}(s;1)$ and $V_{\alpha,n}(s;0)$ are increasing functions of $v$. Thus, from  (\ref{eq:11}), $V_{\alpha,n}(s)$ is an increasing function of $v$.
\end{Proof}

\begin{corollary}\label{corr:1}
For $\alpha>0$, $s_{1}=(v,0)$ and $\bar{s}_{1} = (v+x,0)$, the following relation holds,
\begin{align}
    V_{\alpha}(\bar{s}_{1}) - V_{\alpha}(s_{1}) \geq x.
\end{align}
Thus, $V_{\alpha}(s)$ is a strictly increasing function of $v$.
\end{corollary}

\begin{theorem}\label{th:1}
    There exists a threshold policy $\pi$ which is optimal for (\ref{eq:7}).
\end{theorem}

For a threshold policy $\pi$, we denote the threshold with $v_{th}$. Similarly, we denote an optimal threshold for an optimal policy with $v_{th}^{*}$. 

To prove the next results, we extend our state space so that the age of incorrect information can take values from the set of positive real numbers. We use a similar approach as of \cite{banerjee2024preempt,banerjee2024wiopt}. Thus, we consider the following state space,
\begin{align}\label{eq:nn31}
    \bar{\mathcal{S}} = \{(v,0):v\in{\mathbb{R}^{+}}\}\cup \{0,1\}.
\end{align}
Now, under action $a\in\{0,1\}$, we have to carefully choose the transition probabilities, so that the evolution of states is restricted to a countable state space. Specifically, we follow the similar transition law as of (\ref{eq:6}), where $v\in{\mathcal{R}^{+}}$, and it will ensure that if we start from a state $s\in{\bar{\mathcal{S}}}$, its evolution is restricted to a countable state space. For example, let us assume, we start with a state $s=(3.5,0)$. If the immediate action is $a=1$, then the next state will be either $s_{4}$ or $s_{5}$, and from then onward, the evolution will be restricted to $\mathcal{S}$. If the immediate action is $0$, then the next state will be $(4.5,0)$ or $s_{5}$. Thus, we see that the evolution of the state $s=(3.5,0)$ is restricted to the set $\mathcal{S}_{1}$,
\begin{align}
    \mathcal{S}_{1} = \{(3.5+x,0)| x\in{\mathbb{N}}\}\cup \mathcal{S}.
\end{align}

We denote the discounted value function corresponding to the state space $\bar{\mathcal{S}}$, with $\bar{V}_{\alpha}(s)$, $s\in{\bar{\mathcal{S}}}$. As the evolution of the states under any policy is restricted to a countable state set, similar to (\ref{eq:13}), we have,
\begin{align}\label{eq:nn33}
    \bar{V}_{\alpha}(s) = \lim_{n\rightarrow\infty} \bar{V}_{\alpha,n}(s),
\end{align}
where $\bar{V}_{\alpha,n}(s)$ follows a similar iteration as of (\ref{eq:12}). Similarly, we consider a bounded state space,
\begin{align}
    \bar{\mathcal{S}}_{K} = \{(v,0):v\in{\mathbb{R}^{+},v\leq K}\}\cup \{0,1\}.
\end{align}
Similar to (\ref{eq:nn15}), (\ref{eq:nn16}), we define the relative value functions on $\bar{\mathcal{S}}_{K}$, as $\bar{V}_{n}(s)$ and $\bar{V}(s)$, where $\bar{V}_{n}(s)$ follows a similar iteration as of (\ref{eq:nn15}), and $\bar{V}(s)$ follows a similar relation as of (\ref{eq:nn16}). Similar to (\ref{eq:nn33}), we have,
\begin{align}
    \lim_{n\rightarrow\infty}  \bar{V}_{n}(s) = \bar{V}(s),
\end{align}
for $s\in{\bar{\mathcal{S}}}_{K}$. Note that,  if $s\in{\mathcal{S}}_{K}\cap\bar{\mathcal{S}}_{K}$, then
\begin{align}
    \bar{V}(s) = V(s).
\end{align}
In the rest of the paper, for simplicity, we remove the sub-script $K$ from $\bar{\mathcal{S}}_{K}$ and $\mathcal{S}_{K}$, as we do not use (\ref{eq:3}) and (\ref{eq:nn31}), in the rest of the paper.

Now we define a function $f:\mathbb{R}^{+}\rightarrow \mathbb{R}$ as, 
\begin{align}
    f(v) = \bar{V}((v,0)).
\end{align}
Using a similar technique as of \cite{banerjee2024preempt, banerjee2024wiopt}, we show that $f(v)$ is a concave function of $v$. Thus, from \cite{bertsekas2003convex}, we say that the left-sided derivative of $f$ exists. Now, we mention a remark, which we will use to prove later theorems. 
\begin{remark}\label{rem}
If $f(x)$ is an increasing concave function of $x$, then the following relation holds true,
\begin{align}
    \int_{x_{1}}^{x_{2}} \pdv{f(x)^{-}} {x} dx = f(x_{2}) - f(x_{1}).
\end{align}
\end{remark}

In the next lemma, we find an upper bound and a lower bound on the left-sided derivative, which we will use to prove later results.
\begin{lemma}\label{lemma:2}
     The left-sided derivative of $f$ with respect to $x$, at $v$, i.e.,  $\pdv{f(x)^{-}} {x}|_{v}$ is upper bounded by $\frac{S}{q}$. Similarly,  $\pdv{f(x)^{-}} {x}|_{v}$ is lower bounded by $S$.
\end{lemma}

\begin{Proof}
    From the definition of the left-sided derivative we have,
    \begin{align}
        \pdv{f(x)^{-}} {x} = \lim_{h \rightarrow 0+} \frac{f(x)-f(x-h)}{h}.
    \end{align}
    Now, we can show that we can find a small enough positive $h$, such that an action is optimal for all the real numbers in the interval $(x-h,x)$. Now, assume that for the state $(x,0)$, action $a=0$ is optimal. Then, 
     \begin{align}
        \pdv{f(x)^{-}} {x} =& \lim_{h \rightarrow 0+} \frac{f(x;0)-f(x-h;0)}{h},\label{eq:n27} \\ =& S + \lim_{h \rightarrow 0+} \frac{(1-q)(f(x+1)-f(x-h+1))}{h}, \\ = & S+ (1-q) \pdv{f(x+1)^{-}} {x}.
    \end{align}
    Now, assume that for the state $(x,0)$, action $a=1$ is optimal. Then,
    \begin{align}
        \pdv{f(x)^{-}} {x} =& \lim_{h \rightarrow 0+} \frac{f(x;1)-f(x-h;1)}{h}, \\ = & S.\label{eq:n31}
    \end{align}
    Now, if action $a=0$ is optimal for state $(x,0)$, then from Theorem~\ref{th:1}, we say that there exists a $y\in{\mathbb{R}^{+}}$, such that $(x+y,0)$ is the first state where $a=1$ is optimal, and any state $(x+z,0)$ action $a=1$ is optimal, where $z\geq y$. Thus, from (\ref{eq:n27}), we have,
    \begin{align}
          \pdv{f(x)^{-}} {x} =& \sum_{i=0}^{y-1} S (1-q)^{i} + (1-q)^{y} \pdv{f(x+y)^{-}} {x}, \\ = & \sum_{i=0}^{y-1} S (1-q)^{i} + (1-q)^{y} \pdv{f(x+y;1)^{-}} {x},\label{eq:n33}\\ =&\sum_{i=0}^{y} S (1-q)^{i},\label{eq:n34} \\ \leq & \sum_{i=0}^{\infty} S (1-q)^{i} = \frac{S}{q},
    \end{align}
    where (\ref{eq:n33}) follows from the fact that action $a=1$ is optimal for state $(x+y,0)$, and (\ref{eq:n34}) follows from (\ref{eq:n31}). 

    From (\ref{eq:n34}), we have,
    \begin{align}
         \pdv{f(x)^{-}} {x} =& \sum_{i=0}^{y} S (1-q)^{i}, \\ \geq& S,  
    \end{align}
    which proves this lemma.
\end{Proof}

Now, we study some relations, which will be useful to prove later theorems. 

\begin{lemma}\label{lemma:3}
    The following relations hold,
    \begin{align}
        &\bar{V}(s_{3}) = \frac{1+q}{q} \bar{V}(s_{5}),\label{eq:30} \\ & \bar{V}(s_{4}) = \left(\frac{1}{q} 
    +1 -\frac{1}{q_{1}}\right)  \bar{V}(s_{5}).\label{eq:31}
    \end{align}
\end{lemma}

In the next theorem, we study a sufficient condition on the age of incorrect information, for $a=1$ to be an optimal action.

\begin{theorem}\label{th:thr}
    For a state $s=(v,0)\in{\bar{\mathcal{S}}}$, if $v\geq {\frac{p q}{S(1-q)}}$, then action $a=1$ is optimal for state $s$. If $s\in{\mathcal{S}}$, and $v\geq \ceil{\frac{p q}{S(1-q)}}$, then $a=1$ is optimal for state $s$
\end{theorem}

\begin{Proof}
    From the statement of this theorem, for state $s=(v,0)$, we have the following,
    \begin{align}
        &\frac{p q}{1-q} \leq  Sv, \\  &p q \leq  (1-q)\int_{1}^{v+1}{S} dx , \\ &pq \leq (1-q) \int_{1}^{v+1}  \pdv{f(x)^{-}} {x} dx, \\ &pq \leq (1-q) (f(v+1)-f(1)), \label{eq:n41}
    \end{align}
    From Lemma~\ref{lemma:3}, we have,
    \begin{align}
        \bar{V}(s_{3}) -\bar{V}(s_{4}) = \frac{\bar{V}(s_{5})}{q_{1}}.
    \end{align}
    Thus,
    \begin{align}\label{eq:n43}
        \bar{V}(s_{2}) -\bar{V}(s_{3}) = \bar{V}(s_{2}) -\bar{V}(s_{4}) - \frac{\bar{V}(s_{5})}{q_{1}}.
    \end{align}
    As $\bar{V}(s_{5})$ is the optimal cost corresponding to an optimal policy, $\bar{V}(s_{5})\geq 0$. Thus, from (\ref{eq:n43}), we have,
    \begin{align}
        \bar{V}(s_{2}) - \bar{V}(s_{3}) \leq \bar{V}(s_{2}) -\bar{V}(s_{4}).
    \end{align}
    Thus, from (\ref{eq:n41}), we have,
    \begin{align}
        &pq \leq (1-q) \bar{V}(s_{2}) -\bar{V}(s_{4}), \\ &\bar{V}(s_{1};1) \leq \bar{V}(s_{1};0),  
    \end{align}
    which proves this theorem.
\end{Proof}
Note that, if $S\geq p$ and $q\leq \frac{1}{2}$, always sampling policy is optimal.
Next, we study a necessary condition on the age of incorrect information for action $a=1$ to be optimal. The proof for the next theorem is similar to that for Theorem~\ref{th:thr}.

\begin{theorem}\label{th:3}
    A necessary condition that action $a=1$ is optimal for state $s=(v,0)\in{\mathcal{S}}$, according to the Bellman equation is,
    \begin{align}
        v \geq \max\left\{{\ceil{\frac{pq^{2}}{(1-q)S} -\frac{1}{2q_{1}}},1}\right\}.
    \end{align}
\end{theorem}

We denote a sampling policy, which has a threshold structure with a threshold $v_{th}\in\mathbb{N}$, with $\pi^{v_{th}}$. According to the definition of a threshold policy, if the age of incorrect information is strictly less than $v_{th}$, the sampler chooses action $a=0$, otherwise, it chooses action $a=1$. 

Now, we find the average cost for policy $\pi^{v_{th}}$. Note that the evolution of states under the policy $\pi^{v_{th}}$ is restricted to the following set,
\begin{align}
    \mathcal{S}_{1} = \{(0,0), (0,1), (1,0), (2,0),\cdots, (v_{th},0)\}.
\end{align}
For simplicity, we enumerate the states as follows,
\begin{align}
    &(0,0)=0, \  (0,1)=1, \ (1,0) = 2, \ \cdots, \ (v_{th},0) = v_{th}+1.
\end{align}
Now we find the stationary distribution of the Markov chain $M^{th}$, induced by the policy $\pi^{v_{th}}$, on the state space $\mathcal{S}_{1}$. We denote the stationary distribution of $M^{th}$, with $\bm{p}^{th}$. Note that, the stationary distribution should follow the following equations,
\begin{align}
     &p^{th}(0)\! =\!  p^{th}(0) (1-q) \!+\! q (p^{th}(2)\!+\! \cdots \!+\! p^{th}(v_{th}+1)),\nonumber \\ &p^{th}(1) = (1-q) p^{th}(1) + (1-q) p^{th}(v_{th}+1),\nonumber \\ &p^{th}(2) = q p^{th}(0) + q_{1} p^{th}(1),\nonumber \\ &p^{th}(3) = (1-q) p^{th}(2) ,\nonumber \\ & \vdots \nonumber \\&p^{th}(v_{th}+1) 
     =(1-q) p^{th}(v_{th}).\label{eq:27}
\end{align}
Solving the set of linear equation in (\ref{eq:27}), we get,
\begin{align}
     p^{th}(2) = \frac{q_{1}q} {2 q_{1} - 2 q_{1}(1-q)^{v_{th}} + q (1-q)^{v_{th}}}.
\end{align}
Note that, the cost incurred by the policy $\pi^{th}$ is,
\begin{align}\label{eq:nn58}
     \sum_{i=0}^{v_{th}-1} S (i+1) p^{th}(2) (1-q)^{i} + q (1-q)^{v_{th}-1} p^{th}(2) p.
\end{align}
We summarize this result in the following theorem.
\begin{theorem}\label{th:n4}
     The average cost for a threshold policy $\pi^{v_{th}}$, with the threshold $v_{th}$ is,
     \begin{align}
         \sum_{i=0}^{v_{th}-1} S  (i+1) p^{th}(2) (1-q)^{i} + q (1-q)^{v_{th}-1} p^{th}(2) p,\nonumber
     \end{align}
     where
     \begin{align}
        p^{th}(2) = \frac{q_{1}q} {2 q_{1} - 2 q_{1}(1-q)^{v_{th}} + q (1-q)^{v_{th}}}.\nonumber
    \end{align} 
\end{theorem}

\begin{figure}[t]
    \centerline{\includegraphics[width = 0.9\columnwidth]{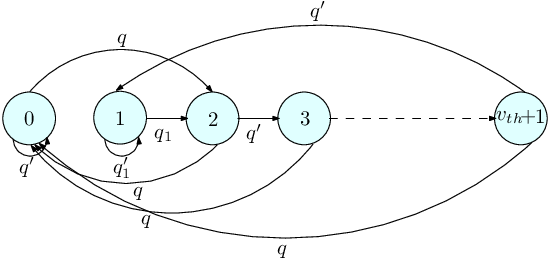}}
    \caption{A pictorial representation of $M^{th}$. Here, we assume that $q'=1-q$ and $q_{1}'=1-q_{1}$. }
    \label{fig:3}
    \vspace*{-0.4cm}
\end{figure}

Thus, from Theorem~\ref{th:thr}, Theorem~\ref{th:3}, and Theorem~\ref{th:n4}, we see that to find $v_{th}^{*}$, i.e., to find an optimal policy, we have to search in the set $\{\max\left\{{\ceil{\frac{pq^{2}}{(1-q)S} -\frac{1}{2q_{1}}},1}\right\},2,\cdots,\ceil{\frac{q p}{s}}\}$. Thus, we can choose $K$ to be as large as possible without changing the numerical complexity. 

Next, we show how the optimal threshold $v_{th}^{*}$ behaves with varying parameters of the problem. 

\begin{theorem}\label{th:5}
    The optimal threshold $v_{th}^{*}$ does not decrease with increasing $p$.
\end{theorem}

\begin{Proof}
     For simplicity of the presentation of this proof, we denote the continuous relative value function at state $(v,0)$, for a penalty $p$, with
     \begin{align}
         &\bar{V}((v,0)) = f(v;p), \qquad \bar{V}((v,0);0) = f(v;0,p),\nonumber\\
         & \bar{V}((v,0);1) = f(v;1,p).
     \end{align}
    We prove this theorem by contradiction. Let us assume that if we increase $p$, the optimal threshold $v_{th}^{*}$ decreases. For notational convenience we write the optimal threshold as $v_{th,p}^{*}$, to demonstrate the dependence on $p$. Consider a penalty for dropping a job, $p_{1}$. Thus, if the age of incorrect information is strictly less than $v_{th,p_{1}}^{*}$, then action $a=0$ is optimal, otherwise action $a=1$ is optimal. From Theorem~\ref{th:thr}, we know that $v_{th,p_{1}}^{*}$ is finite. As $f(v;0,p_{1})$ and $f(v;1,p_{1})$ are both continuous functions of $v$, we have,
    \begin{align}\label{eq:65}
         f(v_{th,p_{1}}^{*};0,p_{1}) = f(v_{th,p_{1}}^{*};1,p_{1}).
    \end{align}
    As in Corollary~\ref{corr:1}, we can show that $f(v;p)$ is a strictly increasing function of $v$. Similarly, we can show that $f(v;0,p)-f(v;1,p)$ is a strictly increasing function of $v$. Thus, we have the following relations for $\epsilon>0$,
    \begin{align}
        &f(v_{th,p_{1}}^{*}+\epsilon;1,p_{1}) < f(v_{th,p_{1}}^{*}+\epsilon;0,p_{1}), \\ & f(v_{th,p_{1}}^{*}-\epsilon;0,p_{1}) < f(v_{th,p_{1}}^{*}-\epsilon;1,p_{1}).\label{eq:67}
    \end{align}
    Now, let us consider another penalty, $p_{2}>p_{1}$. According to our assumption, we say that, 
    \begin{align}
        v_{th,p_{2}}^{*} < v_{th,p_{1}}^{*}.
    \end{align}
    Thus, according to (\ref{eq:67}), the following should hold,
    \begin{align}\label{eq:69}
    f(v_{th,p_{1}}^{*};1,p_{2}) < f(v_{th,p_{1}}^{*};0,p_{2}).
    \end{align}
    As $v_{th,p2}^{*}\leq v_{th,p1}^{*}$, from (\ref{eq:n34}), for $x\in{\mathbb{R}^{+}}$, we say that,
    \begin{align}
        \pdv{f(x;p_{2})^{-}} {x} \leq \pdv{f(x;p_{1})^{-}} {x}.
    \end{align}
    Thus,
    \begin{align}
        \int_{0}^{1} \pdv{f(x;p_{2})^{-}} {x} dx\leq& \int_{0}^{1}\pdv{f(x;p_{1})^{-}} {x} dx, \\  f(1;p_{2}) -f(0;p_{2}) \leq& f(1;p_{1}) - f(0;p_{1}),\label{eq:72}\\ {f(0;p_{2})} \leq & {f(0;p_{1})}, \label{eq:73}
    \end{align}
    Similarly,
    \begin{align}
        \int_{1}^{v+1} \pdv{f(x;p_{2})^{-}} {x} dx\leq& \int_{1}^{v+1}\pdv{f(x;p_{1})^{-}} {x} dx,\\f(v+1;p_{2}) -f(1;p_{2}) \leq& f(v+1;p_{1}) - f(1;p_{1}), \label{eq:75}
    \end{align}
    where (\ref{eq:72}) follows from Remark~\ref{rem} and (\ref{eq:73}) follows from Lemma~\ref{lemma:3}. 
    Now,
    \begin{align}
        \bar{V}(s_{1};0) & - \bar{V}(s_{1};1) \nonumber\\
        =& (1-q)(\bar{V}(s_{2})-\bar{V}(s_{4})) - pq,\\ 
        =& (1-q)(\bar{V}(s_{2})-\bar{V}(s_{3})) + \frac{1-q}{q_{1}} \bar{V}(s_{5})-pq.
    \end{align}
    Thus,
    \begin{align}
        &f(v_{th,p}^{*};0,p) - f(v_{th,p}^{*};1,p) = (1-q)\nonumber\\
        &(f(v_{th,p}^{*}+1;p) - f(1;p)) +\frac{1-q}{q_{1}} f(0;p) -pq. \label{eq:78}
    \end{align}
    As $p_{2}>p_{1}$, from (\ref{eq:73}) and (\ref{eq:75}), we have,
    \begin{align}\label{eq:79}
     &f(v_{th,p_{1}}^{*};0,p_{2}) -  f(v_{th,p_{1}}^{*};1,p_{2})  \nonumber\\
     & \qquad < f(v_{th,p_{1}}^{*};0,p_{1}) - f(v_{th,p_{1}}^{*};1,p_{1}).
    \end{align}
    From (\ref{eq:65}), we have,
    \begin{align}
    f(v_{th,p_{1}}^{*};0,p_{1}) - f(v_{th,p_{1}}^{*};1,p_{1}) =0.
    \end{align}
    Thus,  from (\ref{eq:79}), we have,
    \begin{align}
    f(v_{th,p_{2}}^{*};0,p_{2}) < f(v_{th,p_{2}}^{*};1,p_{2}),
    \end{align}
    which contradicts (\ref{eq:69}).
\end{Proof}

The proofs for Theorem~\ref{th:4}, Theorem~\ref{th:n5} and Theorem~\ref{th:7} are similar to the proof for Theorem~\ref{th:5}.

\begin{theorem}\label{th:4}
   The optimal threshold $v_{th}^{*}$ does not increase with increasing $S$.
\end{theorem}
 
\begin{theorem}\label{th:n5}
     The optimal threshold $v_{th}^{*}$ does not decrease with increasing $q_{1}$.
\end{theorem}
  
\begin{theorem}\label{th:7}
The optimal threshold $v_{th}^{*}$ does not decrease with increasing $q$, when $q\geq \frac{1}{2}$.
\end{theorem}

\begin{figure}[t]
    \centerline{\includegraphics[width = 1\columnwidth]{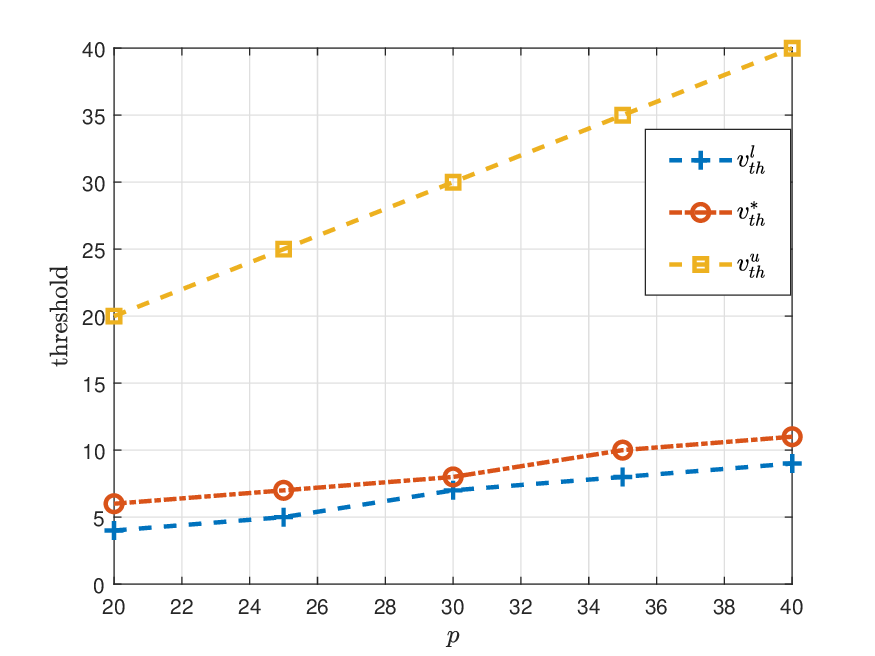}}
    \caption{We compare the optimal threshold, a lower and an upper bound of optimal threshold for different $p$.}
    \label{fig:simu1}
\end{figure}

\begin{figure}[t]
    \centerline{\includegraphics[width = 1\columnwidth]{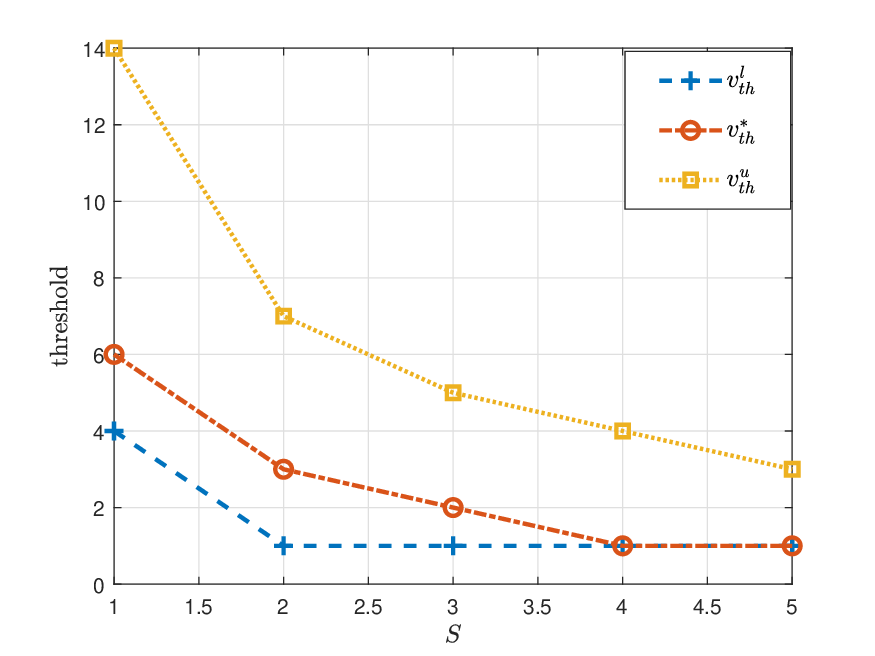}}
    \caption{We compare the optimal threshold, a lower and an upper bound of optimal threshold for different $S$.}
    \label{fig:simu2}
\end{figure}

\begin{figure}[t]
    \centerline{\includegraphics[width = 1\columnwidth]{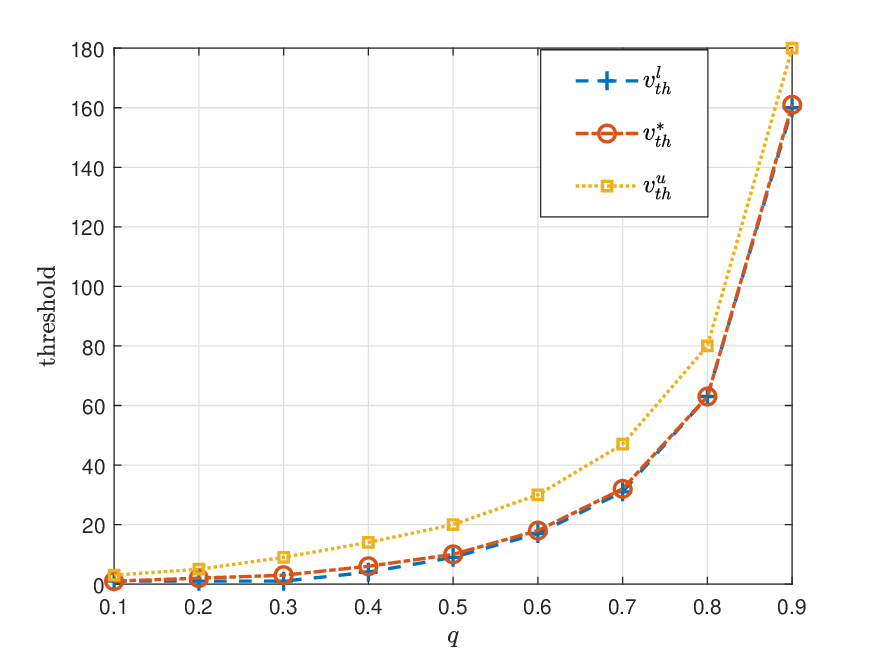}}
    \caption{We compare the optimal threshold, a lower and an upper bound of optimal threshold for different $q$.}
    \label{fig:simu3}
\end{figure}

\section{Numerical Results}
\begin{figure}[t]
    \centerline{\includegraphics[width = 1\columnwidth]{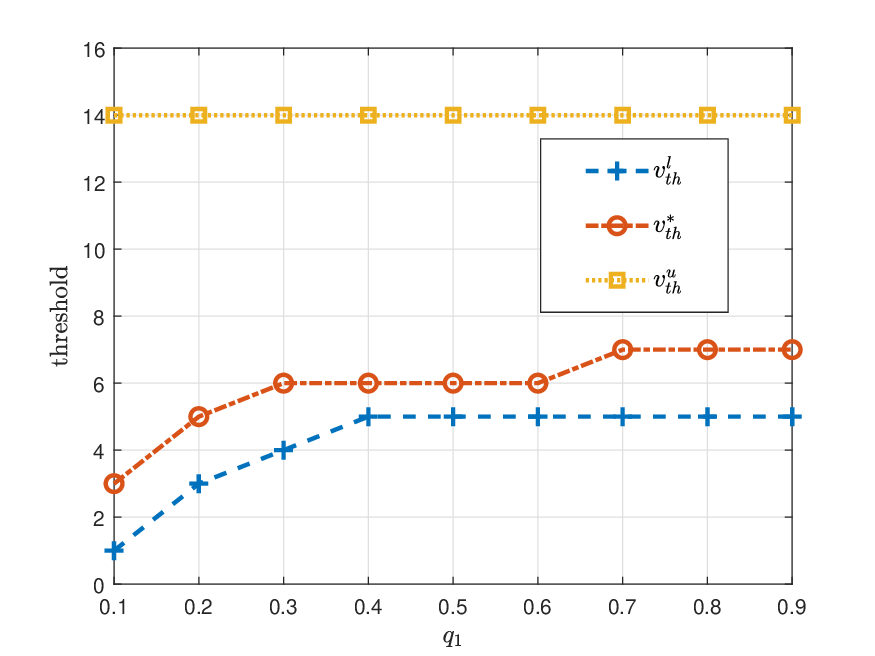}}
    \caption{We compare the optimal threshold, a lower and an upper bound of optimal threshold for different $q_{1}$.}
    \label{fig:simu4}
\end{figure}

\begin{figure}[t]
    \centerline{\includegraphics[width = 1\columnwidth]{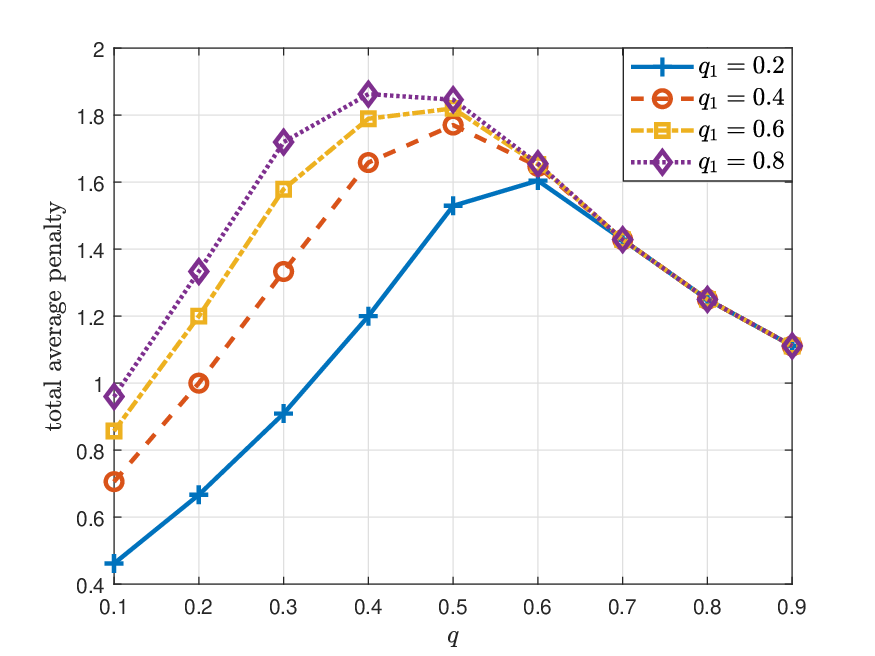}}
    \caption{We compare the total average penalty corresponding to the optimal threshold policy for varying $q$ and $q_{1}$.}
    \label{fig:simu5}
\end{figure}

First, we verify the results of Theorem~\ref{th:5}, Theorem~\ref{th:4}, Theorem~\ref{th:n5} and Theorem~\ref{th:7}, in Fig.~\ref{fig:simu1}, Fig.~\ref{fig:simu2}, Fig.~\ref{fig:simu3}, and Fig.~\ref{fig:simu4}, respectively. For these figures, we denote the optimal threshold with $v_{th}^{*}$, the lower bound on the optimal threshold corresponding to Theorem~\ref{th:3} with $v_{th}^{l}$, and the upper bound on the optimal threshold corresponding to Theorem~\ref{th:thr} with $v_{th}^{u}$. For Fig.~\ref{fig:simu1}, we consider $S=1$, $q=0.4$, and $q_{1}=0.3$. We see from Fig.~\ref{fig:simu1} that the optimal threshold increases with penalty $p$, verifying the Theorem~\ref{th:5}. For Fig.~\ref{fig:simu2}, we consider $p=20$, $q=0.4$ and $q_{1}=0.3$. We see from Fig.~\ref{fig:simu2} that the optimal threshold decreases with $S$, which validates Theorem~\ref{th:4}. For Fig.~\ref{fig:simu3}, we consider $S=1$, $p=20$ and $q_{1}=0.3$. From Fig.~\ref{fig:simu3}, we see that thee optimal threshold increases with $q$, validating Theorem~\ref{th:n5}. For Fig.~\ref{fig:simu4}, we consider $S=1$, $P=20$ and $q=0.4$. From Fig.~\ref{fig:simu4}, it is immediate that the statement of Theorem~\ref{th:7} holds. Note that, the rate at which the optimal threshold increases with $q$ is much larger than the rate at which the optimal threshold increases with $q_{1}$. In Fig.~\ref{fig:simu5}, we study how the average penalty under the optimal policy changes with $q$ and $q_{1}$. We see that for a fixed $q$, the optimal total average penalty increases with $q_{1}$. For a fixed $q_{1}$, the optimal penalty first increases and then decreases with increasing $q$. We also notice that the $q$ at which the optimal penalty starts decreasing decreases with increasing $q_{1}$.

\bibliographystyle{unsrt}
\bibliography{references}

\end{document}